\documentclass[10pt,a4paper]{article}
\usepackage[utf8]{inputenc}
\usepackage{geometry}
\usepackage{graphicx}
\usepackage{amsmath,amssymb}
\usepackage{dsfont}
\usepackage{color}
\usepackage{amsmath}
\usepackage{verbatim} 

\begin{document}

\begin{titlepage}
\begin{flushright}
October 2020
\end{flushright}
\begin{flushright}
\end{flushright}
\vfill
\begin{center}
{\Large\bf Spontaneous Symmetry Breaking}
\vfill
{\bf Andreas Aste}\\[1cm]
{\it Department of Physics, University of Basel,\\
Klingelbergstrasse 82, \\
CH-4056 Basel, Switzerland\\
E-Mail: andreas.aste@unibas.ch}
\end{center}
\vfill
\begin{abstract}
The concept of spontaneous symmetry breaking (SSB) generally lacks
a simple and intuitive introduction in the literature. This gap is filled by defining SSB
in a universal context beyond its usual applications in physics and by discussing
some very simple, but stunning examples of the phenomenon. 
\end{abstract}
\vfill
\vfill
\end{titlepage}

\section{Introduction}
A symmetry is a map(ping) of a structured object $\mathcal{O}$ onto itself which preserves the structure.
For example, if $\mathcal{O}$ is a set with no additional structure, a symmetry is a bijective map
from the set onto itself. If the object $\mathcal{O}$ is a set of points in the Euclidean plane,
a symmetry is a bijection of the set onto itself which preserves the distance between each pair of points
in the plane.

In most cases of interest, the structured object under study is a structured set. Due to the many intricacies of
the mathematical definition of an object, we will restrict us to the well-established notions of
conventional set theory. Within this context, one should keep in mind that the elements of an object
can be non-trivially structured objects as well.

The set $S(\mathcal{O})$ of all symmetries of a structured set $\mathcal{O}$ forms the symmetry group
$(S(\mathcal{O}), \circ)$ through the group law given by the composition $\circ$ of symmetries:
if $s_1$, $s_2 \in S(\mathcal{O})$, then $s_1 \circ s_2 \in S (\mathcal{O})$, where
\begin{equation}
s_1: \mathcal{O} \leftrightarrow \mathcal{O} \, , \quad
s_2: \mathcal{O} \leftrightarrow \mathcal{O} \, , \quad
(s_1 \circ s_2) (x) = s_1(s_2(x)) \, \, \forall \, x \in \mathcal{O} \, .
\end{equation}

Every structured object possesses the trivial symmetry given by the identity map $n \in S (\mathcal{O})$
which leaves all elements of $\mathcal{O}$ invariant: $n(x)=x \, \, \forall x \in \mathcal{O}$.

Given a mathematical or physical problem, one has to distinguish between the problem and its solutions,
which both represent structured objects.
E.g., consider the problem to find the shortest transport network consisting of line segments
between four cities located at the vertices of a (unit) square \cite{Raifeartaigh:1993}.
The symmetry group of this problem is isomorphic to the dihedral group $D_4$,
if reflection symmetries are allowed, or the cyclic group $C_4$ otherwise.

One might be tempted to assume that a solution of a problem possesses the symmetries
of the problem itself, i.e., $D_4$ in the present case.
A first guess for an optimized network as shown in Figure 1 on the left has a total
length of $\sqrt{8}$ and the full $D_4$-symmetry, but it does not represent a solution to the posed problem.
In fact, there are exactly two solutions also shown in Figure 1 with a total length of the
transportation network given by $\sqrt{3}+1 \simeq 2.73205 \ldots$,
having a smaller symmetry group $D_2 < D_4$. The symmetry of the {\emph{problem}}
is {\emph{spontaneously}} broken by the {\emph{solutions}}.
However, rotating the $1^{st}$
solution by an angle of $\pi / 4 = 90^o$ leads to the $2^{nd}$ solution. This fact expresses
the symmetry of the problem itself; applying a symmetry transformation of the problem to a solution of
the problem again leads to a solution.\\

In the following, the simple fact that a solution has a smaller symmetry group than its constituting problem
will be referred to as SSB in the {\emph{general}} sense. The complete absence of {\emph{any}}
solution with the full symmetry of its constituting problem will be referred to as SSB in the {\emph{usual}}
or {\emph{narrow}} sense.
$\quad$\\
\vskip 0.3cm
\begin{minipage}[c]{3.5cm}
\includegraphics[scale=0.15]{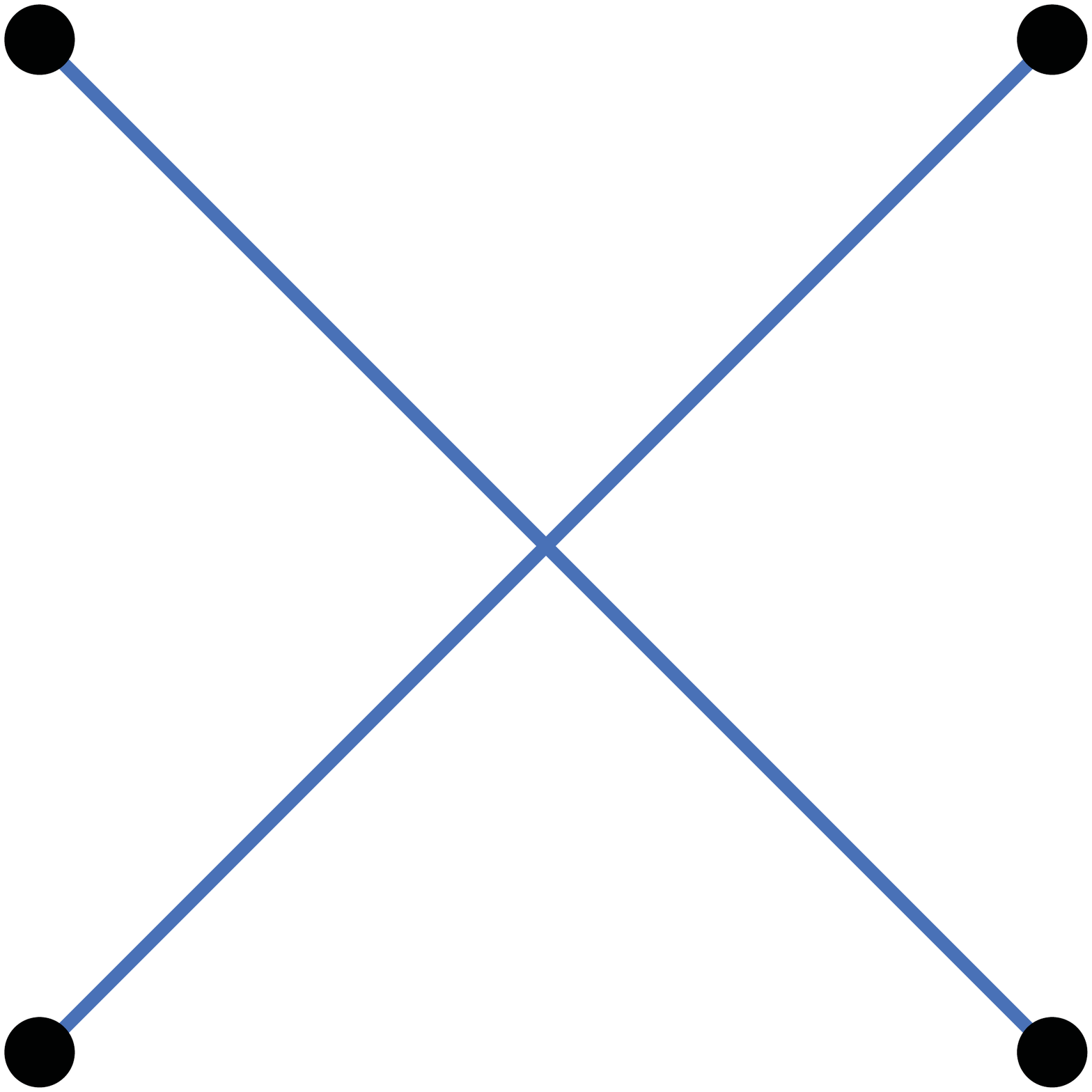}\\
\vskip 0.1cm \hskip 1.25cm
{\textcolor{blue}{Not optimal}}
\end{minipage}
\begin{minipage}[c]{2.1cm}
$ $
\end{minipage}
\begin{minipage}[c]{3.5cm}
\begin{center}
\vskip -0.73cm
\includegraphics[scale=0.1525]{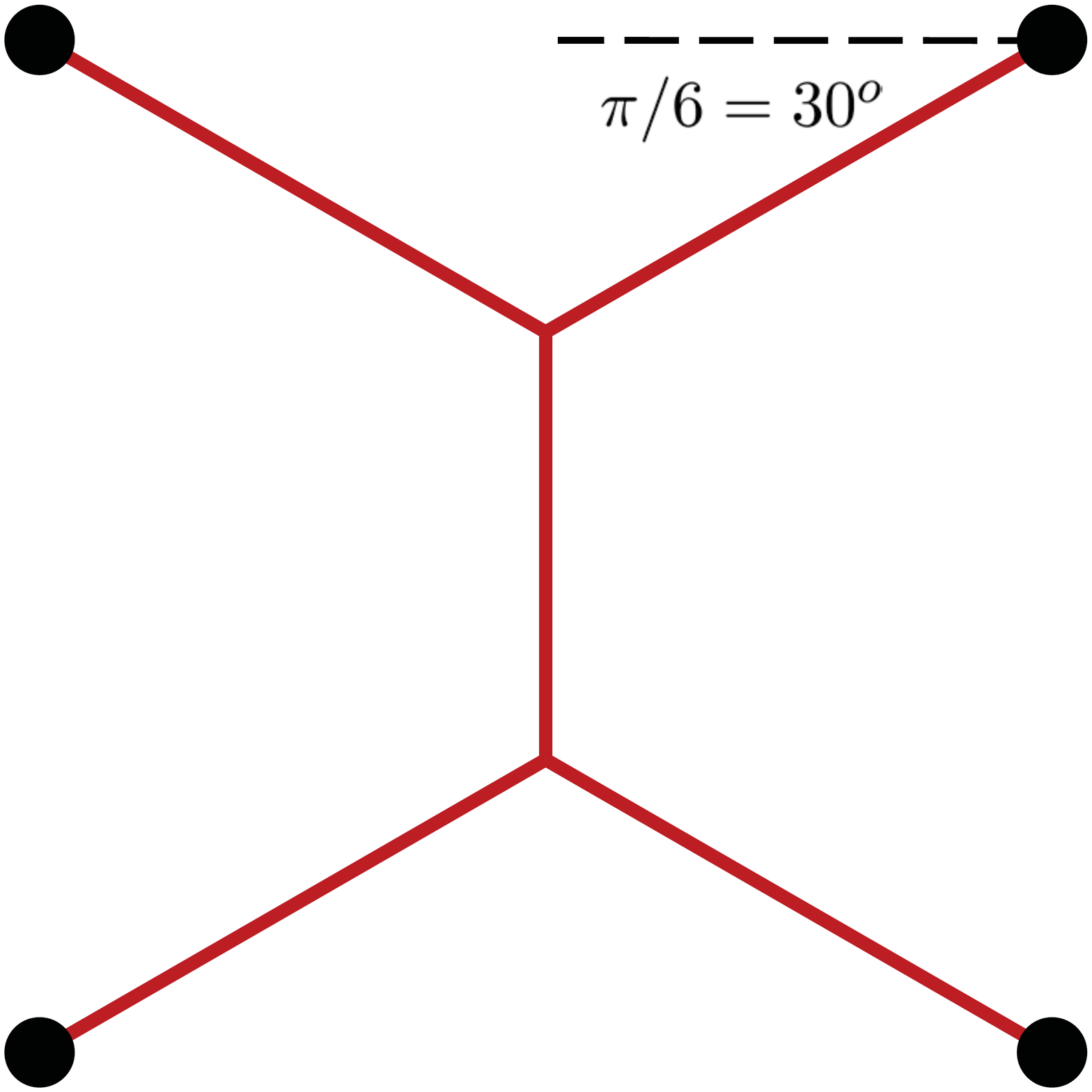}\\
\vskip -0.135cm
{\textcolor{red}{$1^{st}$ solution}}
\end{center}
\end{minipage}
\begin{minipage}[c]{0.3cm}
$ $
\end{minipage}
\begin{minipage}[c]{3.5cm}
\vskip -0.1cm
\begin{center}
\includegraphics[scale=0.1525, angle=90]{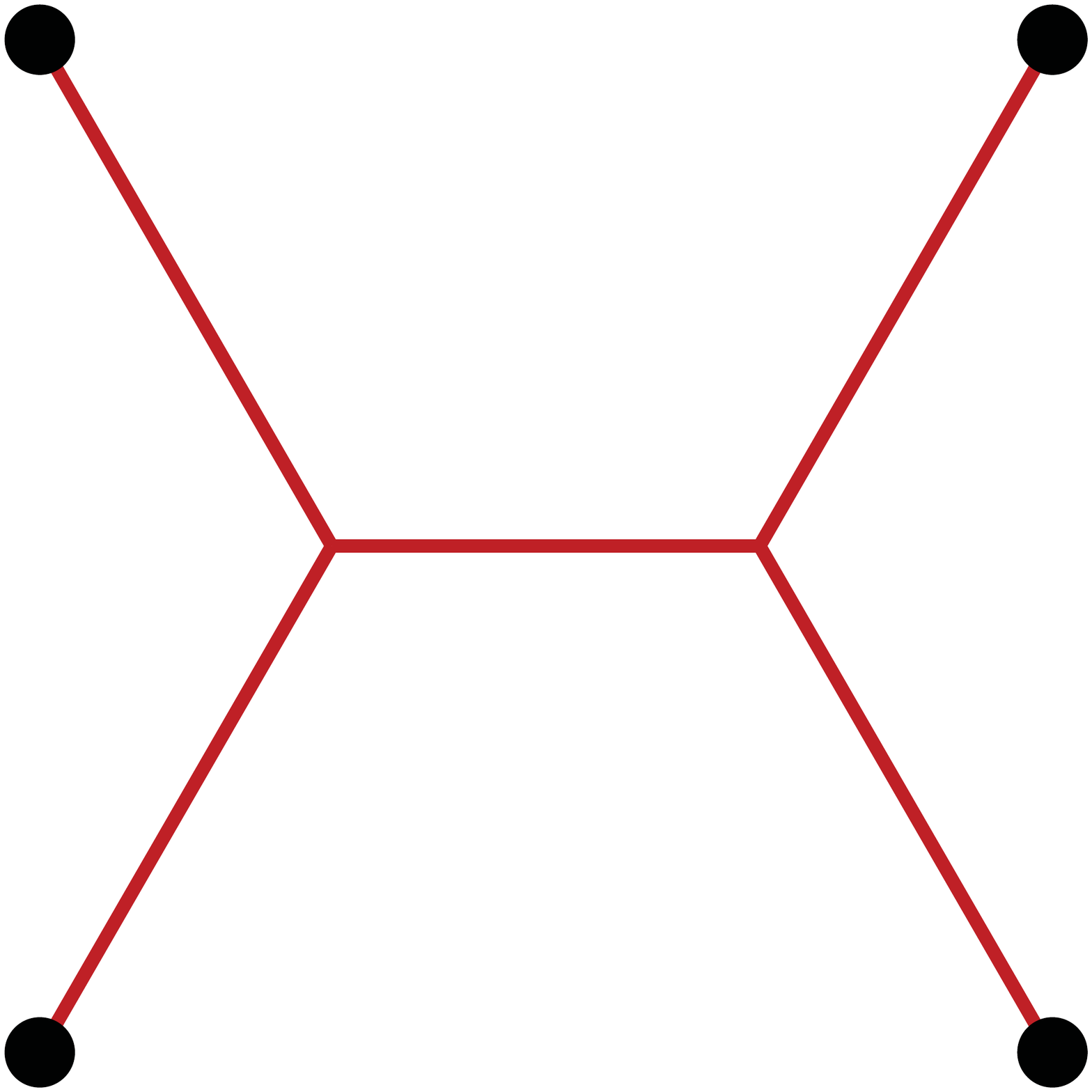}\\
\vskip 0.5cm \hskip 1.05cm
{\textcolor{red}{$2^{nd}$ solution}}
\end{center}
\end{minipage}\\
\vskip 0.4cm
Figure 1: The shortest networks connecting the vertices of a square do not possess the
symmetry $D_4$ of the square: the total length of an intuitive $D_4$-symmetric guess
is $\sqrt{8} = 2.82842 \ldots $ for a unit square, whereas the total length of the
two shortest networks with symmetry group $D_2$ is
$\sqrt{3}+1 \simeq 2.73205 \ldots$.

\section{General formulation}
Most problems in mathematics or physics can be formulated in the following way:
A problem can be associated (or identified) with an operator $\mathcal{P}$, and one
has to look for a set of solutions $\mathcal{S} \subseteq \mathcal{M}$ within a more
general set $\mathcal{M}$ of potential solutions, such that $\mathcal{P} [ f ] = 0$ when
the problem is solved by some $f \in \mathcal{S}$ or $\mathcal{P} [ f ] \neq 0$ holds otherwise.

\begin{itemize}
\item A very simple example is given by the following $\mathcal{P}_1$: Solve the real equation
\begin{equation}
x^2 = 1 \, , \quad \mbox{or} \quad \mathcal{P}_1 [x] = x^2 -1 = 0 \, ,
\end{equation}
which obviously has the symmetries $x \rightarrow x$ and $x \rightarrow -x$, and therefore
the symmetry group of $\mathcal{P}_1$ is isomorphic to $\mathds{Z}_2$.
However, the solutions $\mathcal{S}_1 = \{ \pm 1 \} \subset \mathcal{M}_1 = \mathds{R}$
change sign under the symmetry transformation,
and one is lead to a similar situation as observed in the problem presented in the
introduction. Selected symmetries of the problem connect the different solutions, SSB in the narrow sense
occurs, since no fully $\mathds{Z}_2$-symmetric solution exists.

The solutions of the problem $\tilde{\mathcal{P}}_1 [x] = -x^2 +x^4 = x^2 (x^2-1) =0$ are
$\tilde{\mathcal{S}}_1 = \{ \pm 1, 0 \}$. In this modified $\mathds{Z}_2$-symmetric problem,
a $\mathds{Z}_2$-symmetric solution exists, namely $x=0$.

A subtle phenomenological difference appears when one looks for the position of the {\emph{minima}}
of the polynomial $p(x) = -x^2+x^4$ by solving first the problem $\tilde{\mathcal{P}}' [x] = 4 x^3 -2 x = 0$.
$p(x)$ has two minima at $x_{1,2} = \pm 1/\sqrt{2}$ and a local maximum at $x_0 = 0$.
The local maximum is $\mathds{Z}_2$-symmetric, but it is not a valid solution of the actual problem.
The analogy of this situation with the Higgs mechanism is obvious. $x_{1,2}$ are the non-symmetric
`stable vacua' of the problem, SSB in the narrow sense occurs, since $x_0=0$ is excluded from the set
of solutions due to its `instability' as a local maximum.

\item A less trivial problem $\mathcal{P}_2$ consists in finding the solutions of the real differential equation
\begin{equation}
f'(x) = f(x) \, , \quad \mbox{or} \quad
\mathcal{P}_2 [ f ] = \biggl( \frac{d}{dx} -1 \biggr) f \equiv 0 \, ,
\end{equation}
which has translational symmetry in the sense that the equation
\begin{equation}
f'(x+a) = f(x+a) \, , \quad a \in \mathds{R}
\end{equation}
represents the same problem, when we chose $\mathcal{M}_2 = C^1 (\mathds{R} )$,
the set of all continuously differentiable real functions on the real line.
The set of solutions is
$\mathcal{S}_2 = \{ f_c \mid f_c (x) = c e^x , \, c \in \mathds{R} \} \subset
\mathcal{M}_2$.
Again, the functions $f_c$ do {\emph{not}} exhibit translational symmetry like the
$\mathcal{P}_2$-operator itself, one has
\begin{equation}
f_c (x+a)  \neq f_c (x) \quad \mbox{for} \quad c \neq 0 \, .
\end{equation}
{\emph{However}}, there {\emph{is}} a solution with the full symmetry of $\mathcal{P}_2$,
namely $f_0 (x) \equiv 0$.

$f_0$ is a solution of the highest symmetry, i.e. the symmetry of the problem itself,
which can be associated in a physical sense with the notion of a {\emph{vacuum}} state,
since expectation values of observables on such a symmetric state usually take on extremal values. The
absence of such a solution or state would represent the phenomenon of SSB on a deeper level
(in the narrow sense), in contrast to the SSB in a general sense displayed by the exponential
functions $f_{c \neq 0}$ expressing the fact that solutions of a problem often are less symmetric than
the underlying problem itself. In the physical context, a vacuum solution is required to be
stable, as it happens to be the case in the following example.

\item In order to study the symmetry properties of a system of partial differential
equations and its solutions, one may consider Maxwell's equations {\emph{in vacuo}}
for the electric field $E$ and the magnetic field $B$ in compact notation
\begin{equation}
\nabla \cdot E = \nabla \cdot B =0 \, , \quad
\nabla \times E = - \dot{B} \, , \quad \nabla \times B = \dot{E} \, . \label{maxvac}
\end{equation}
The replacement
\begin{equation}
E' = \lambda E - \lambda' B \, , \quad B'= \lambda  B + \lambda' E \, , \quad
\lambda, \lambda' \in \mathds{R} \label{rotscal}
\end{equation}
leads to an equivalent set of equations for $E'$ and $B'$.
Introducing the photon wave function \cite{Aste:2012dc}
\begin{equation}
F = E + iB \,  , \quad i^2 = -1 \, ,
\end{equation}
Maxwell's equations in vacuo read $\nabla \cdot F =0$ and $\dot{F} = -i \nabla \times F$, and
the transformations (\ref{rotscal}) can be written in the form
\begin{equation}
F' = z F \quad \mbox{with} \quad z = \lambda + i \lambda' \in \mathds{C}_{\neq 0} =
\mathds{C} \backslash \{ 0 \} \, ,
\end{equation}
hence the symmetry group of the equations is isomorphic to the multiplicative group of
complex numbers $(\mathds{C}_{\neq 0} , \cdot)$.

There is a solution to Maxwell's equations (\ref{maxvac}) in vacuo which is invariant
under the full symmetry group given by the transformations (\ref{rotscal}), namely
the electromagnetic vacuum field state $F \equiv 0$. A non-vanishing field configuration has a smaller symmetry group than
Maxwell's equations, but $F \equiv 0$ remains unchanged by a multiplication with $z \in \mathds{C}_{\neq 0}$.
\end{itemize}

To conclude, a less trivial example for spontaneous symmetry breaking shall be investigated in the
following section.

\section{Scaling symmetry and spontaneous breaking of scale invariance
in an unphysical gauge sector}
A function $f(\vec{x})$ depending on $\vec{x} = (x_1, \ldots , x_n) \in \mathds{R}^n$
is scale-invariant, when the essential properties of the function remain unchanged
under a rescaling $\vec{x} \mapsto \lambda \vec{x}$.
This requirement is usually interpreted in the sense that the function $f$ changes only by a factor
$s(\lambda)$ when submitted to a scaling transformation according to
\begin{equation}
f( \lambda \vec{x})=s(\lambda) f(\vec{x}) \, , \quad \lambda \neq 0 \, .
\end{equation}

E.g., the electrostatic Coulomb potential $\Phi (\vec{x})$ of a point charge $q$ in 3-dimensional Euclidean space
is invariant under scaling transformations $s_\lambda$ with $\lambda \neq 0$
and $r = || \vec{x} || = (x_1^2+x_2^2+x_3^2)^{1/2}$ according to
\begin{equation}
\Phi (\vec{x}) = \frac{q}{4 \pi \epsilon_0} \frac{1}{r} \quad \rightarrow \quad \Phi' (\vec{x}) =
(s_\lambda \Phi ) (\vec{x}) = \lambda \Phi (\lambda \vec{x}) =
\frac{\lambda q}{4 \pi \epsilon_0} \frac{1}{\lambda r} =\Phi (\vec{x}) \, .
\end{equation}
With the neutral element $s_1$ and the group law given by the composition
$s_{\lambda_1} \circ s_{\lambda_2} = s_{\lambda_1 \lambda_2}$, the symmetry group
of $\Phi$ is isomorphic to the multiplicative group of the real numbers
\begin{equation}
S(\Phi) = (\{ s_\lambda \mid \lambda \in \mathds{R}  \backslash \{ 0 \} \}, \circ )
\simeq ( \mathds{R}_{\neq 0}, \cdot ) \, .
\end{equation}

Turning now to electrostatics in $\mathds{N} \ni n$-dimensional Euclidean space, the Poisson equation
for the electrostatic potential $\Phi$ of a point charge located in the space origin reads
\begin{equation}
\Delta \Phi(\vec{x})=-q \delta^{(n)}(\vec{x}) \, , \quad \vec{x} \in \mathds{R}^n \, ,
\label{Poisson_dim_n}
\end{equation}
where, e.g., for $n=3,4$ the spherically symmetric solutions vanishing for $r \rightarrow \infty$
(where the radial distance from the origin is now $r = || \vec{x}|| = (x_1^2 + \ldots x_n^2)^{1/2}$) are
\begin{align}
\displaystyle
n=3: \quad & \vec{E} (\vec{x}) =-\vec{\nabla} \Phi  (\vec{x})
= \frac{q}{4 \pi} \frac{\vec{x}}{r^3} \, , \quad \quad \, \, \,
\Phi  (\vec{x}) = \frac{q}{4 \pi} \frac{1}{r} \, , \\
n=4: \quad & \vec{E}  (\vec{x}) =-\vec{\nabla} \Phi  (\vec{x}) =
\frac{q}{2 \pi^2} \frac{\vec{x}}{r^4} \, , \quad \quad
\Phi  (\vec{x}) = \frac{q}{4 \pi^2} \frac{1}{r^2} \, ,
\end{align}
and in the general case for $n \in \mathds{N}$ and $n > 2$ one has
\begin{equation}
\displaystyle
\vec{E} (\vec{x}) =-\vec{\nabla} \Phi  (\vec{x})
= \frac{q}{O_{n-1}} \frac{\vec{x}}{r^n}, \quad
\Phi  (\vec{x}) = \frac{q}{(n-2) O_{n-1}} \frac{1}{r^{n-2}} \, ,
\quad O_{n-1} =\frac{2 \pi^{n/2}}{\Gamma (n/2)} \, , \label{Gauss_n_dim}
\end{equation}
where $O_{n-1}$ is the
$(n-1)$-dimensional volume of the $(n-1)$-dimensional unit sphere
$S^{n-1} \subset \mathds{R}^n$.
The equations (\ref{Gauss_n_dim}) are a consequence of Gauss's flux theorem,
stating that the electric flux through a closed surface enclosing any volume is the total charge enclosed within
that volume (expressed above with natural units where $\epsilon_0 = 1$).

Due to the scaling properties of the Laplace operator and the Dirac delta distribution
$(\lambda > 0$)
\begin{equation}
\Delta \Phi(\lambda \vec{x})=\lambda^2 (\Delta \Phi)(\lambda \vec{x}) \quad
\mbox{und} \quad
\delta^{(n)}(\lambda \vec{x})=\lambda^{-n} \delta^{(n)} (\vec{x})
\label{Skalen_Del_Del}
\end{equation}
one might expect for the spherically symmetric solutions of the Poisson equation
(\ref{Poisson_dim_n}) that the corresponding general scaling symmetry
\begin{equation}
\Phi(\lambda \vec{x})=\frac{1}{\lambda^{n-2}} \Phi(\vec{x})
\label{Skalen_Phi}
\end{equation}
also holds for the electrostatic potentials given by equations (\ref{Gauss_n_dim}),
and this is in fact the case for $n>2$.

By a change of variables
$\vec{x} \rightarrow \lambda \vec{x}$, the Poisson equation
(\ref{Poisson_dim_n}) reads
\begin{equation}
(\Delta \Phi) (\lambda \vec{x}) = -q \delta^{(n)} (\lambda \vec{x}) \, ,
\end{equation}
and due to the scaling symmetries given by the equations (\ref{Skalen_Del_Del})
then follows
\begin{equation}
\lambda^{-2} \Delta ( \Phi (\lambda \vec{x})) = - \lambda^{-n} q \delta^{(n)} (\vec{x}) \, .
\label{Poisson_rescal}
\end{equation}
Assuming that the scaling property (\ref{Skalen_Phi}) indeed holds for $\Phi$,
equation (\ref{Poisson_rescal}) becomes
\begin{equation}
\lambda^{-2} \lambda^{-n+2} \Delta \Phi (\vec{x}) = - \lambda^{-n} q 
\delta^{(n)} (\vec{x}) \, ,
\end{equation}
and therefore a rescaled field $\Phi'(\vec{x}) = \lambda^{n-2} \Phi (\lambda \vec{x})$
is a solution again as a consequence of the scaling symmetry of the Poisson equation:
\begin{equation}
\Delta \Phi' (\vec{x}) = \lambda^n
\underbrace{( \Delta \Phi) (\lambda \vec{x})}_{- q \delta^{(n)} (\lambda \vec{x})}
= -q \delta^{(n)} ( \vec{x} ) \, .
\end{equation}

Now, in two dimensions something unexpected happens:
one straightforwardly calculates
\begin{eqnarray}
n=2: \quad E  (\vec{x}) =\frac{q}{2 \pi} \frac{1}{r}, \quad
\Phi_\mu  (\vec{x}) = - \frac{q}{2 \pi} \log (r/\mu). 
\end{eqnarray}
The scaling symmetry of the two-dimensional Poisson equation for a point charge
is broken by the solutions $\Phi_\mu$ with the highest possible spherical symmetry,
and a length scale $\mu$ appears spontaneously, also corresponding to an energy- or
mass scale from a physical point of view.

One observes that spherically symmetric solutions of the Poisson equation do {\emph{not}}
exhibit the full symmetry of their constituting Poisson equation (\ref{Poisson_dim_n}) for
$n=2$; however, when $\Phi_\mu (\vec{x})$ is a solution, then a rescaled field
\begin{equation}
\Phi_\mu (\lambda \vec{x})= \Phi_{\mu/\lambda} (\vec{x})=
\Phi_\mu (\vec{x}) - (q / 2 \pi) \log \lambda
\end{equation}
is also a solution, an this new solution differs from the original one by a scale invariant
{\emph{gauge constant}} $\sim \log \lambda$.
Therefore, the SSB observed in two dimensions only affects the {\emph{unphysical gauge
sector}} of our model theory, since the electrostatic potential $\Phi$ does not correspond
to an observable quantity in the fictitious two-dimensional world. This feature is rather common
in the framework of SSB in (quantum) gauge theories. The electric field
$\vec{E} (\vec{x}) = q \vec{x}/ ( 2 \pi r^2)$,
however, corresponds to a measurable gauge invariant quantity (in QED, but not in QCD),
and it still shows the regular scaling behavior for $n=2$ also observed in other dimensions
$n \neq 2$.\\

The appearance of the scale parameter can be related to a regularization procedure in
the space of tempered distributions $\mathcal{S}(\mathds{R}^2)$ (containing the
Dirac delta distribution as an element), which may serve as
the set $\mathcal{M}$ of potential solutions to the two-dimensional Poisson equation with point charge.

\section{Conclusions}
Spontaneous symmetry breaking can be viewed, in a general and rather trivial sense, as the simple fact
that the solution of a given problem has a smaller symmetry group than the problem itself.
Whereas the natural laws in flat spacetime are Poincar\'e invariant, i.e., invariant with respect
to spacetime translations and proper orthochronous Lorentz transformations, even a single particle
state governed by these laws has a smaller symmetry group. The cosmic microwave background
is not Lorentz invariant as well as a human body.

In a narrow sense usually adopted as a viewpoint in field theories,
spontaneous symmetry breaking describes the absence of specific solutions,
often associated with {\emph{stable}} vacuum states in physics, which share the full
symmetry group of the underlying principles governing the structure of the system under study.
Then SSB can be unobservable, when it affects an unphysical sector of the involved theory only.

\end{document}